\documentclass[aip,jcp,reprint,showkeys]{revtex4-1}
\usepackage{bm,graphicx,tabularx,array,dcolumn,xcolor,microtype,multirow,amscd,amsmath,amssymb,amsfonts,physics,siunitx}
\usepackage[version=4]{mhchem}
\usepackage[utf8]{inputenc}
\usepackage[T1]{fontenc}
\usepackage{txfonts}
\usepackage{simpler-wick}

\usepackage{dblfloatfix,placeins,float}

\usepackage{lipsum}

\usepackage[
	colorlinks=true,
    citecolor=blue,
    linkcolor=blue,
    filecolor=blue,      
    urlcolor=blue,	
    breaklinks=true
	]{hyperref}
\usepackage{caption} 
\usepackage{subcaption}

\newcommand{\cF}{\mathcal{F}}
\newcommand{\cJ}{\mathcal{J}}

\newcommand{\cV}{\mathcal{V}}

\newcommand{\up}[1]{\hspace{0pt}^{#1}}
\newcommand{\Or}[1]{\mathcal{O}(#1)}
\newcommand{\LibGNME}{\textsc{LibGNME}}

\newcommand{\Ne}{N}
\newcommand{\Nbas}{n}
\newcommand{\Nex}{L}

\newcommand{\vdag}{\vphantom{\dagger}}
\newcommand{\tS}{\tilde{S}}

\newcommand{\etal}{\textit{et al.}}
\newcommand{\eg}{e.g.}
\newcommand{\ie}{i.e.}
\definecolor{hughgreen}{RGB}{0, 139, 0}

\newcommand{\UOX}{Physical and Theoretical Chemistry Laboratory, Department of Chemistry, University of Oxford, South Parks Road, Oxford, OX1 3QZ, U.K.}

\begin{document}

\title{Generalized nonorthogonal matrix elements for arbitrary excitations}
\author{Hugh~G.~A.~Burton}
\email{hgaburton@gmail.com}
\affiliation{\UOX}

\date{\today}

\begin{abstract}
Electronic structure methods that exploit nonorthogonal Slater determinants face the
challenge of efficiently computing  nonorthogonal matrix elements.
In a recent publication, 
\href{https://doi.org/10.1063/5.0045442}{H.\ G.\ A.\ Burton, J.\ Chem.\ Phys.\ 154, 144109 (2021)}, 
I introduced a
generalized nonorthogonal extension to Wick's theorem that allows matrix elements to be derived between
excited configurations from different reference determinants.
However, that work only provided explicit expressions for one- and two-body matrix elements between singly-
or doubly-excited configurations.
Here, this framework is extended to compute generalized nonorthogonal matrix elements between arbitrary excitations.
Pre-computing and storing intermediate values allows one- and two-body matrix elements to be evaluated with an $\Or{1}$ 
scaling relative to the system size, and the \LibGNME{} computational library 
is introduced to achieve this in practice.
These advances make the evaluation of nonorthogonal matrix elements almost as easy as their orthogonal 
counterparts, facilitating a new phase of development in nonorthogonal electronic structure theory.
\end{abstract}

\maketitle
\raggedbottom
\linepenalty1000
\twocolumngrid

\section{Introduction}

Nonorthogonality occurs throughout modern quantum chemistry, providing chemically intuitive and compact
representations of challenging electronic structure.
For example, nonorthogonal configuration interaction (NOCI) uses a linear combination of
nonorthogonal Slater determinants, each with bespoke orbitals representing an important electronic 
configuration, to capture strong static correlation%
\cite{Koch1993,TenNo1997,Ayala1998,Thom2009,Sundstrom2014,Mayhall2014,Burton2019c,Huynh2019,Burton2020,Lee2022} 
or provide a diabatic representation of electron transfer states.\cite{Jensen2018,Wibowo2017,Straatsma2022} 
Alternatively, nonorthogonal transition coupling elements arise for orbitally-optimized excited states,
which can improve the accuracy of challenging core\cite{Oosterbaan2018,Oosterbaan2019} 
or charge transfer excitations.\cite{Gilbert2008,Hait2020,CarterFenk2020,Levi2020,Shea2018,Hardikar2020,Tran2019,Tran2020}
Beyond providing chemical intuition, nonorthogonality naturally arises 
in auxiliary-field quantum Monte-Carlo,\cite{Mahajan2020,Mahajan2021}
in coupling terms between states at different molecular geometries,\cite{Chen2022,SLee2021} 
and in spin-projected techniques.\cite{Scuseria2011,Tsuchimochi2016,Tsuchimochi2016b,Lee2022}

Despite presenting many potential benefits, the development of nonorthogonal electronic
structure methods is hindered by the difficulty of computing matrix elements between Slater determinants
with mutually nonorthogonal orbitals.
For orthogonal orbitals, the second-quantized generalized Wick's theorem\cite{ShavittBook} allows arbitrary 
matrix elements to be easily derived and evaluated using pre-computed one- and two-electron integrals.
In contrast, matrix elements for nonorthogonal determinants currently require
the first-quantized generalized Slater--Condon rules,\cite{MayerBook} with a computational scaling of 
at least $\Or{\Ne^3}$ with respect to the number of electrons $\Ne$.\cite{Thom2009}

While the generalized Slater--Condon rules  can be used to couple a small number of Slater determinants,%
\cite{Thom2009,Burton2016,Burton2019c,Sundstrom2014,Mayhall2014} 
their additional computational cost quickly makes larger calculations unfeasible.
In particular, many recent developments involve multiple orthogonal excited 
configurations from different nonorthogonal reference determinants, introducing a large number 
of nonorthogonal matrix elements.
This situation arises for the coupling terms between state-specific 
multi-determinant wave functions\cite{Shea2018,Hardikar2020,Tran2019,Tran2020,Kossoski2022}
or post-NOCI techniques for capturing dynamic correlation,\cite{TenNo1997,Tsuchimochi2016,Tsuchimochi2016b} 
including perturbative approximations such as 
NOCI-MP2\cite{Yost2013,Yost2016,Yost2019} and NOCI-PT2.\cite{Burton2020}
Developing efficient implementations of these methods requires a theory for computing nonorthogonal
matrix elements between excited configurations that does not scale with the system size.
While equations for specific applications have been independently derived many times 
(e.g.\ Refs.~\onlinecite{Sundstrom2014} and \onlinecite{Yost2013}), 
an entirely generalized approach has not yet been developed.
In this work, I derive a theory that achieves this aim for arbitrary one- and 
two-body nonorthogonal matrix elements.

The first significant step towards an entirely general theory for nonorthogonal matrix elements 
was the nonorthogonal extension to Wick's theorem.\cite{Hendekovic1981}
However, that approach is restricted to reference determinants with a non-zero 
many-body overlap, excluding the case where the orbitals are mutually nonorthogonal but the 
determinants themselves have a zero overlap,\cite{RingBook} and has seen limited use in quantum
chemistry.\cite{Jimenez-Hoyos2012a,Tsuchimochi2016,Tsuchimochi2016b}
Similarly, Mahajan and Sharma introduced an approach for efficiently evaluating Hamiltonian
matrix elements between an arbitrary multi-determinant wave function and a single reference determinant
built from different orbitals.\cite{Mahajan2020,Mahajan2021}
While promising, this theory also fails for orthogonal reference determinants 
with mutually nonorthogonal orbitals.

In Ref.~\onlinecite{Burton2021a} (henceforth known as Paper 1), 
I introduced a generalized nonorthogonal extension 
to Wick's theorem that allows matrix elements to be computed 
for any pair of nonorthogonal determinants, even if their many-body overlap is zero.
This theory allows arbitrary overlap, one-body, and two-body coupling terms to be derived 
for excited configurations from different reference determinants, 
and enables a second-quantized derivation of the generalized Slater--Condon rules.
I then demonstrated that overlap and one-body coupling terms between 
nonorthogonal excited configurations can be evaluated with $\Or{1}$ scaling
using a set of pre-computed intermediates.
However, these explicit derivations were difficult to extend for coupling terms beyond double excitations 
and the $\Or{1}$ scaling did not appear to be achievable for two-electron coupling terms.

In the current work, I extend the results of Paper~1 by introducing a simplified 
framework for computing nonorthogonal matrix elements between arbitrary excited configurations
from different reference determinants.
This framework combines the extended nonorthogonal Wick's theorem derived in Paper~1 with 
a determinantal expansion of Wick's theorem based on L\"{o}wdin's general 
formula,\cite{Lowdin1955a,Lowdin1955b} giving universal expressions for arbitrary excitations.
Furthermore, by introducing pre-computed intermediates, 
I derive explicit expressions for the overlap, one-body, and two-body coupling terms 
between arbitrary excitations which scale as $\mathcal{O}(1)$ with respect to the 
number of electrons or basis functions.
These advances significantly reduce the computational cost compared to the previous
state-of-the-art, creating a new paradigm for efficient nonorthogonal matrix elements 
in quantum chemistry.

In Section~\ref{sec:BackgroundTheory}, I introduce the key results from Paper~1, before 
extending these to arbitrary excitations in Section~\ref{sec:extension}.
Section~\ref{sec:results} presents numerical data illustrating the computational scaling of 
these generalized nonorthogonal matrix elements.
The primary conclusions and scope for future applications are summarized in Section~\ref{sec:conc}.

\section{Background Theory}
\label{sec:BackgroundTheory}

In this Section, I summarize the key aspects of the extended nonorthogonal Wick's theorem 
derived in Paper 1 that are required for the derivations presented in this work. 
Interested readers are directed to Paper 1 for a comprehensive explanation of these results
and to Ref.~\onlinecite{ShavittBook} for a detailed description of the generalized Wick's theorem.

\subsection{Biorthogonalising the molecular orbitals}
This work considers matrix elements between excited configurations constructed 
from a general pair of mutually nonorthogonal determinants, \eg{}
\begin{equation}
\begin{split}
&\mel*{\up{x}\Phi_{ij\dots}^{ab\dots}}{\hat{O}}{\up{w}\Phi_{kl\dots}^{cd\dots}} 
=
\\
&\quad\qquad\mel{\up{x}\Phi}{
(\up{x}b_i^{\dagger}\up{x}b_j^{\dagger} \cdots \up{x}b_b\up{x}b_a)
\hat{O}
(\up{w}b_c^{\dagger}\up{w}b_d^{\dagger} \cdots \up{w}b_l\up{w}b_k)
}{\up{w}\Phi}  
\end{split}
\label{eq:genMatEl}
\end{equation}
where $\hat{O}$ is an arbitrary operator.
Here, $\ket{\up{x}\Phi}$ and $\ket{\up{w}\Phi}$ are reference Slater determinants built from different molecular
orbitals $\{\up{x}\phi_p\}$ and $\{\up{w}\phi_p\}$ that are expanded using a common $\Nbas$-dimensional 
atomic orbital basis set $\{\chi_\mu \}$ as
\begin{align}
\up{x}\phi_p &= \sum_{\mu}^{\Nbas} \chi_{\mu}\, \up{x}C^{\mu \cdot}_{\cdot p}
\quad\text{and}\quad
\up{w}\phi_p = \sum_{\mu}^{\Nbas} \chi_{\mu}\, \up{w}C^{\mu \cdot}_{\cdot p}.
\end{align}
The $\up{x}C^{\mu \cdot}_{\cdot p}$ denote the orbital coefficients in the molecular orbital basis of 
determinant $\ket{\up{x}\Phi}$ using the nonorthogonal tensor notation of Head-Gordon \etal{}\cite{HeadGordon1998}
The corresponding second-quantized operators $\up{x}b^{\dagger}_p$  and $\up{x}b_p$ create and
annihilate, respectively, an electron in the molecular orbital $p$ for the orbital set 
corresponding to determinant $\ket{\up{x}\Phi}$.

To evaluate matrix elements using these nonorthogonal determinants, the molecular orbitals
$\{\up{x}\phi_p\}$ and $\{\up{w}\phi_p\}$ must be transformed into a biorthogonal set 
using the L\"{o}wdin pairing approach.\cite{Amos1961,Hall1951} 
First, the overlap matrix between the two sets of occupied orbitals is computed as
\begin{equation}
\up{xw}S_{ij} = \sum_{\mu \nu}^{\Nbas} (\up{x}C^{*})^{\cdot \mu}_{i \cdot} g_{\mu \nu} (\up{w}C)^{\nu \cdot}_{\cdot j},
\end{equation}
where $g_{\mu \nu} = \braket*{\chi_\mu}{\chi_\nu}$ is the overlap matrix for the atomic orbital basis set.
A singular value decomposition is then performed to diagonalize this overlap matrix, giving the modified
occupied orbital coefficients $(\up{w}\tilde{C})^{\cdot \mu}_{i \cdot}$ and 
$(\up{x}\tilde{C})^{\cdot \mu}_{i \cdot}$ that satisfy
\begin{equation}
\sum_{\mu \nu}^{\Nbas}  
(\up{x}\tilde{C}^{*})^{\cdot \mu}_{i \cdot} g_{\mu \nu} (\up{w}\tilde{C})^{\nu \cdot}_{\cdot j}
=
\up{xw}\tilde{S}_i\, \delta_{ij}.
\end{equation}
When one, or more, of the diagonal overlap terms becomes zero (i.e., $\up{xw}\tilde{S}_i=0$), 
the many-body overlap must vanish $\braket{\up{x}\Phi}{\up{w}\Phi} = 0$.
The number of these biorthogonal zero-overlap orbital pairs corresponds to the 
nullity of the matrix $\up{xw}\bm{S}$ and is denoted by the integer $m$.
The value of $m$ determines which instance of the generalized Slater--Condon rules is 
required for a matrix element between the reference states $\ket{\up{x}\Phi}$ and $\ket{\up{w}\Phi}$, 
as described in Refs.~\onlinecite{MayerBook} and \onlinecite{Thom2009}.
The product of the remaining non-zero singular values defines the reduced overlap
\begin{equation}
\up{xw}\tilde{S} = \prod_{\{ i | \up{xw}\tilde{S}_{i} \neq 0\}} \up{xw}\tilde{S}_{i},
\end{equation}
which also appears in the generalized Slater--Condon rules.\cite{Thom2009}

\subsection{Extended nonorthogonal Wick's theorem}
Without loss of generality, I will represent operators in the molecular orbital basis for determinant 
$\ket{\up{x}\Phi}$.
For example, a one-body operator $\hat{f}$ is given by
\begin{equation}
\hat{f} = \sum_{pq} \up{x}f_{pq} \up{x}b_{p}^{\dagger} \up{x}b_{q}
\end{equation}
where 
\begin{equation}
\begin{split}
\up{x}f_{pq} 
= \mel*{\up{x}\phi_p}{\hat{f}}{\up{x}\phi_q}
= \sum_{\mu \nu} (\up{x}C^{*})^{\cdot \mu}_{p \cdot}\,\mel*{\chi_\mu}{\hat{f}}{\chi_\nu}\,(\up{x}C)^{\nu \cdot}_{\cdot q}
.
\end{split}
\end{equation}
A matrix element such as
\begin{equation}
\begin{split}
\langle \up{x}\Phi_{i\cdots}^{a\cdots} | &\hat{f} | \up{w}\Phi_{j\cdots}^{b\cdots} \rangle
\\
&= 
 \sum_{pq} \up{x}f_{pq} \langle \up{x}\Phi  |
\up{x}b_{i}^{\dagger} \up{x}b_{a} 
\cdots
\up{x}b_{p}^{\dagger} \up{x}b_{q} 
\cdots
\up{w}b_{b}^{\dagger} \up{w}b_{j} 
| \up{w}\Phi \rangle
\end{split}
\label{eq:genmatel}
\end{equation}
can then be evaluated using an extension to Wick's theorem (see Ref.~\onlinecite{ShavittBook}).
In particular, each term in Eq.~\eqref{eq:genmatel} is computed as the sum of all 
possible ways to fully contract the second-quantized operators, for example 
\begin{equation}
\begin{split}
\langle \up{x}\Phi  |
\up{x}b_{i}^{\dagger} \up{x}b_{a} 
\cdots
&\up{x}b_{p}^{\dagger} \up{x}b_{q} 
\cdots
\up{w}b_{b}^{\dagger} \up{w}b_{j} 
| \up{w}\Phi \rangle
\\
&=
\mel*{\up{x}\Phi}{
\wick{
\up{x}\c1 b_{i}^{\dagger} \up{x}\c1 b_{a} 
\cdots
\up{x}\c1 b_{p}^{\dagger} \up{x}\c1 b_{q} 
\cdots
\up{w}\c1 b_{b}^{\dagger} \up{w}\c1 b_{j} 
}
}{\up{w}\Phi}
\\
&+
\mel*{\up{x}\Phi}{
\wick{
\up{x}\c2 b_{i}^{\dagger} \up{x}\c1 b_{a} 
\cdots
\up{x}\c1 b_{p}^{\dagger} \up{x}\c2 b_{q} 
\cdots
\up{w}\c1 b_{b}^{\dagger} \up{w}\c1 b_{j} 
}
}{\up{w}\Phi}
\\
&+
\mel*{\up{x}\Phi}{
\wick{
\up{x}\c2 b_{i}^{\dagger} \up{x}\c1 b_{a} 
\cdots
\up{x}\c1 b_{p}^{\dagger} \up{x}\c1 b_{q} 
\cdots
\up{w}\c1 b_{b}^{\dagger} \up{w}\c2 b_{j} 
}
}{\up{w}\Phi}
\\
&+ \cdots
\end{split}
\end{equation}
The phase of each term is given by $(-1)^{h}$ where $h$ is the number of intersecting contraction lines.

For the extended nonorthogonal Wick's theorem outlined in Paper 1, a
general matrix element in this expansion (\eg{}, 
$\mel*{\up{x}\Phi}{
\wick{
\up{x}\c1 b_{i}^{\dagger} \up{x}\c1 b_{a} 
\cdots
\up{x}\c1 b_{p}^{\dagger} \up{x}\c1 b_{q} 
\cdots
\up{w}\c1 b_{b}^{\dagger} \up{w}\c1 b_{j} 
}
}{\up{w}\Phi}$)
is equal to the reduced overlap $\up{xw}\tilde{S}$ multiplied by the product of 
nonorthogonal contractions
\begin{subequations}
\begin{align}
\wick{\up{x}\c1 b_{p}^{\dagger} \up{w} \c1 b_{q}} &= \up{wx}X_{qp},
\\
\wick{\up{x}\c1 b_{p} \up{w} \c1 b_{q}^{\dagger}}  &= -\, \up{xw}Y_{pq},
\end{align}
\label{eq:nonorthogonalContractions}
\end{subequations}
where $\up{wx}X_{qp}$ and $\up{xw}Y_{qp}$ are intermediate values defined for a 
given pair of determinants [Eq.~\eqref{eq:funPairContract}].
If there are zero-overlap orbital pairs in the biorthogonal basis  ($m>0$), then
a sum must be taken over every possible way to assign the $m$ zeros to the
contractions in each product.
This distribution is denoted by indices $m_k$ assigned to each contraction 
(\ie{}, $\up{wx}X_{qp}^{(m_k)}$ and $\up{xw}Y_{pq}^{(m_k)}$), which take 
values of 0 or 1 and satisfy $\sum_k m_k = m$.
The individual contractions, represented in the original (un-transformed) orbital basis, are then defined as
\begin{subequations}
\begin{align}
 \up{wx}X_{qp}^{(m_k)} 
&=
\begin{cases}
\sum_{\mu \nu \sigma \tau} (\up{w}C^{*})^{\cdot \mu}_{q \cdot}\, g_{\mu \sigma}(\up{xw}M)^{\sigma \tau}\, g_{\tau \nu}\, (\up{x}C)^{\nu \cdot}_{\cdot p} & m_k=0,
\\
\sum_{\mu \nu \sigma \tau} (\up{w}C^{*})^{\cdot \mu}_{q \cdot}\, g_{\mu \sigma}(\up{xw}P)^{\sigma \tau}\, g_{\tau \nu}\, (\up{x}C)^{\nu \cdot}_{\cdot p} & m_k=1,
\end{cases}
\label{eq:funPairContract1}
\\
 \up{xw}Y_{pq}^{(m_k)} 
&=
\begin{cases}
 \up{xw}X_{pq}^{(m_k)} - \up{xw}S_{pq}	  & m_k=0
\\
\up{xw}X_{pq}^{(m_k)}   & m_k=1,
\end{cases}
\label{eq:funPairContract2}
\end{align}
\label{eq:funPairContract}
\end{subequations}
where
\begin{subequations}
\begin{align}
\up{xw}P_{k}^{\sigma \tau} &= (^{w}\tilde{C})^{\sigma \cdot}_{\cdot k} (^{x}\tilde{C}^{*})^{\cdot \tau}_{k \cdot},
\\
\up{xw}W^{\sigma \tau} &= 
\sum_{\{ i | \up{xw}\tilde{S}_{i} \neq 0\} } 
(^{w}\tilde{C})^{\sigma \cdot}_{\cdot i} \frac{1}{\up{xw}\tilde{S}_{i}} (^{x}\tilde{C}^{*})^{\cdot \tau}_{i \cdot}.
\\
\up{xw}P^{\sigma \tau} &= \sum_{\{k| \up{xw}\tilde{S}_{k} = 0\} } \hspace{-2pt}^{xw}P_{k}^{\sigma \tau},
\label{eq:CoDensitySum}
\\
\up{xw}M^{\sigma \tau} &=  \up{xx}P^{\sigma \tau} + \up{xw}P^{\sigma \tau} + \up{xw}W^{\sigma \tau}.
\label{eq:Msum}
\end{align}
\label{eq:CoDensity}
\end{subequations}
\textit{(\textbf{Warning:} for convenience throughout this work, 
the sign of $\up{xw}Y_{pq}$ has been reversed relative to Paper 1.)}

Notably, the contractions in Eq.~\eqref{eq:nonorthogonalContractions} are defined with 
respect to the original orbital coefficients such that the definition of excited configurations
is not affected by the biorthogonal transformation.
The number of zero-overlap orbital pairs corresponds to the underlying biorthogonal basis and there is no
relationship between the $p,q$ and $m_k$ indices.
Furthermore, the $\up{wx}X_{qp}^{(m_k)}$ and $\up{xw}Y_{pq}^{(m_k)}$ intermediates can be evaluated
and stored once for each pair of nonorthogonal reference determinants.

While the biorthogonal orbital coefficients only enter in the definition of the $\up{xw}\bm{M}$ and $\up{xw}\bm{P}$
matrices, the number of biorthogonal zero-overlap orbital pairs affects every matrix element 
through the summation over the allowed combinations of $m_k$ values.
This distribution is essential for recovering the different instances of the generalized Slater--Condon rules.\cite{Burton2021a}
If $m$ is larger than the total number of contractions, then the corresponding matrix element is strictly zero.

In summary, an arbitrary nonorthogonal matrix element between excited configurations can be 
evaluated through the following procedure:
\begin{enumerate}
\item{\label{step:contract2}%
    Assemble all fully contracted combinations of the second-quantized operator and excitations,
    and compute the corresponding phase factors;}
\item{\label{step:sum_m2}%
    For each term, sum every possible way to distribute $m$ zeros among the 
    contractions $\{m_k\}$ such that $\sum_k m_k = m$;}
\item{\label{step:factor2}%
    For every set of $\{m_k\}$ in each term, construct the relevant contribution 
    as a product of fundamental contractions defined in Eqs.~\eqref{eq:nonorthogonalContractions} and \eqref{eq:funPairContract};}
\item{\label{step:overla2}
    Multiply the combined expression by the reduced overlap $\up{xw}\tilde{S}$.}
\end{enumerate}
To demonstrate this process, consider the matrix element 
$\braket*{\up{x}\Phi_{i}^{a}}{\,\up{w}\Phi_{j}^{b}} =  \mel*{\up{x}\Phi}{\up{x}b_{i}^{\dagger}\up{x}b_{a} \up{w}b_{b}^{\dagger}\up{w}b_{j}}{\up{w}\Phi}$.
Applying Step~\ref{step:contract2} gives two contributions
\begin{equation}
\mel*{\up{x}\Phi}{\wick{\up{x}\c1 b_{i}^{\dagger} \up{x}\c1 b_{a} \up{w}\c1 b_{b}^{\dagger}\up{w}\c1 b_{j} } }{\up{w}\Phi}
+
\mel*{\up{x}\Phi}{\wick{\up{x}\c2 b_{i}^{\dagger} \up{x}\c1 b_{a} \up{w}\c1 b_{b}^{\dagger}\up{w}\c2 b_{j} } }{\up{w}\Phi}.
\end{equation}
Each term corresponds to a product of two fundamental contractions with a phase of $+1$.
Taking a sum over the different $m_1$, $m_2$ values satisfying $m_1 + m_2 = m$, 
and multiplying by the reduced overlap $\up{xw}\tilde{S}$, then gives
\begin{equation}
\braket*{\up{x}\Phi_{i}^{a}}{\,\up{w}\Phi_{j}^{b}} = 
\up{xw}\tilde{S} 
\hspace{-0.25em}
\sum_{\substack{m_1, m_2 \\ m_1 + m_2 = m}}
\hspace{-0.15em}
\qty( \up{xx}X_{ai}^{(m_1)}\,\up{ww}X_{jb}^{(m_2)} - \up{wx}X_{ji}^{(m_1)}\,\up{xw}Y_{ab}^{(m_2)} ),
\end{equation}
which can also be represented as the determinant
\begin{equation}
\braket*{\up{x}\Phi_{i}^{a}}{\,\up{w}\Phi_{j}^{b}} = 
\up{xw}\tilde{S} 
\hspace{-0.25em}
\sum_{\substack{m_1, m_2 \\ m_1 + m_2 = m}}
\begin{vmatrix}
\up{xx}X_{ai}^{(m_1)} && \up{xw}Y_{ab}^{(m_2)} 
\\
\up{wx}X_{ji}^{(m_1)} && \up{ww}X_{jb}^{(m_2)}
\end{vmatrix}.
\end{equation}
This matrix element reduces to zero if $m>2$. 
\onecolumngrid

\begin{widetext}
\section{Extension for arbitrary excitations}
\label{sec:extension}

\subsection{Unification with L\"{o}wdin's general formula}
\label{subsec:LowdinGeneralFormula}
Paper 1 demonstrated the explicit derivation of matrix elements for the overlap, one-body, and two-body 
operators between excited configurations up to double excitations using the extended nonorthogonal 
Wick's theorem.\cite{Burton2021a}
However, those derivations are difficult to generalize for higher-order excitations due to the increasingly
large number of fully-contracted terms in the expansion of a matrix element.
A fully generalized approach can be achieved using L\"{o}wdin's general formula,\cite{Lowdin1955a,Lowdin1955b}
which expresses the matrix element for an arbitrary product of creation and annihilation 
operators as a determinant
and automatically includes every fully-contracted term with the correct phase factor.
Since L\"{o}wdin's general formula is only based on the different diagrammatic combinations of contractions 
and their corresponding phases, it can be readily extended to the nonorthogonal contractions defined in 
Eq.~\eqref{eq:nonorthogonalContractions} to give
\begin{equation}
\mel*{\up{x}\Phi}{%
(\up{x}b^{\dagger}_{p} \up{x}b_{q} ) \cdots
(\up{x}b^{\dagger}_{r} \up{x}b_{s} ) \cdots  
(\up{w}b^{\dagger}_{t} \up{w}b_{u} )%
}{\up{w}\Phi}
=
\up{xw}\tilde{S}
\begin{vmatrix}
\wick{\up{x}\c1 b^{\dagger}_{p} \up{x}\c1 b_{q}} & \cdots & - \wick{\up{x}\c1 b_{q} \up{x}\c1 b^{\dagger}_{r} } & \cdots & - \wick{\up{x}\c1 b_{q} \up{w}\c1 b^{\dagger}_{t}  }
\\
\vdots & & \vdots & & \vdots
\\
\wick{\up{x}\c1 b^{\dagger}_{p} \up{x}\c1 b_{s}} & \cdots & \wick{\up{x}\c1 b^{\dagger}_{r} \up{x}\c1 b_{s}} & \cdots & - \wick{\up{x}\c1 b_{s}  \up{w}\c1 b^{\dagger}_{t} }
\\
\vdots & & \vdots & & \vdots
\\
\wick{\up{x}\c1 b^{\dagger}_{p} \up{w}\c1 b_{u}} & \cdots & \wick{\up{x}\c1 b^{\dagger}_{r} \up{w}\c1 b_{u}} & \cdots & \wick{\up{w}\c1 b^{\dagger}_{t} \up{w}\c1 b_{u} }.
\end{vmatrix}.
\label{eq:LowdinFormula}
\end{equation}
Here, terms in the upper triangle of the determinant correspond to contractions where the annihilation operator 
is to the left of the creation operator and are multiplied by a factor of -1 to give the correct 
sign once the determinant is expanded.

In contrast to the conventional form of L\"{o}wdin's general formula, the entire determinant is 
multiplied by the reduced overlap $\up{xw}\tilde{S}$.
Furthermore, the distribution of $m$ zero-overlap biorthogonal orbital pairs 
over products of individual contractions can be achieved
by exploiting the fact that every term in the expansion of a determinant includes one element 
from each column of the corresponding matrix.
Therefore, a summation can be taken over every possible way to distribute the $m$ zero-overlap orbital
pairs among the columns in Eq.~\eqref{eq:LowdinFormula}.
Inserting the nonorthogonal contractions defined in Eqs.~\eqref{eq:nonorthogonalContractions} and 
\eqref{eq:funPairContract} then gives
\begin{equation}
\begin{split}
\mel*{\up{x}\Phi}{%
(\up{x}b^{\dagger}_{p} \up{x}b_{q} ) \cdots
(\up{x}b^{\dagger}_{r} \up{x}b_{s} ) \cdots  
(\up{w}b^{\dagger}_{t} \up{w}b_{u} )%
}{\up{w}\Phi}
=
\up{xw}\tS \hspace{-0.5em}
\sum_{\substack{m_1, \dots, m_{\Nex} \\ m_1 + \dots + m_{\Nex} = m}}
\begin{vmatrix}
\up{xx}X^{(m_1)}_{qp}    & \cdots & \up{xx}{Y}^{(m_{i})}_{qr} & \cdots & \up{xw}{Y}^{(m_\Nex)}_{qt} 
\\[2pt]
\vdots &  & \vdots & \cdots & \vdots 
\\                                              
\up{xx}X^{(m_1)}_{sp}    & \cdots & \up{xx}{X}^{(m_{i})}_{sr} & \cdots & \up{xw}{Y}^{(m_\Nex)}_{st} 
\\[2pt]
\vdots &  & \vdots & \cdots & \vdots 
\\                                              
\up{wx}X^{(m_1)}_{up}    & \cdots & \up{wx}{X}^{(m_{i})}_{ur} & \cdots & \up{ww}{X}^{(m_\Nex)}_{ut} 
\end{vmatrix},
\end{split}
\label{eq:NonorthogonalLowdinFormula}
\end{equation}
where 
$\Nex$ is the total number of 
creation and annihilation pairs in the operator string.
In Eq.~\eqref{eq:NonorthogonalLowdinFormula}, the lower triangle of the matrix (including the diagonal) 
contains ``$X$-type'' matrix elements
while the upper triangle contains ``$Y$-type'' matrix elements.
Reversing the sign of the ``$Y$-type'' terms defined in Eq.~\eqref{eq:funPairContract2} 
relative to Paper 1 ensures that these elements enter the determinant in Eq.~\eqref{eq:NonorthogonalLowdinFormula}
with a positive sign contribution, simplifying subsequent derivations.
This modified representation of the extended nonorthogonal Wick's theorem using 
L\"{o}wdin's general formula now allows 
matrix elements between arbitrary nonorthogonal excited configurations to be evaluated, as
demonstrated in Sections~\ref{subsec:overlap}--\ref{subsec:two-body}.

\subsection{Overlap Elements}
\label{subsec:overlap}

The overlap between a general pair of excited configurations is given by
\begin{equation}
\begin{split}
\braket*{ \up{x}\Phi_{ij\cdots}^{ab\dots} }{\up{w}\Phi_{kl\cdots}^{cd\dots} }
=
\bra{\up{x}\Phi} (\up{x}b_i^\dagger \up{x}b_a^{\vdag}) (\up{x}b_j^\dagger \up{x}b_b^{\vdag} )
\cdots 
(\up{w}b_d^\dagger \up{w}b_l^{\vdag}) (\up{w}b_c^\dagger \up{w}b_k^{\vdag} )\ket{\up{w}\Phi} .
\end{split}
\end{equation}
Using Lowdin's general representation for the extended nonorthogonal Wick's theorem presented in 
Section~\ref{subsec:LowdinGeneralFormula}, these overlap matrix elements can be expanded as
\begin{equation}
\begin{split}
\braket*{\up{x}\Phi_{ij\dots}^{ab\dots}}{\up{w}\Phi_{kl\dots}^{cd\dots}} 
=
\up{xw}\tS \hspace{-0.5em}
\sum_{\substack{m_1, \dots, m_{\Nex} \\ m_1 + \dots + m_{\Nex} = m}}
\begin{vmatrix}
\up{xx}X^{(m_1)}_{ai}   & \up{xx}Y^{(m_2)}_{aj}  & \cdots& \up{xw}{Y}^{(m_{\Nex-1})}_{ad} & \up{xw}{Y}^{(m_\Nex)}_{ac} 
\\[2pt]
\up{xx}X^{(m_1)}_{bi}   & \up{xx}X^{(m_2)}_{bj}  & \cdots& \up{xw}{Y}^{(m_{\Nex-1})}_{bd} & \up{xw}{Y}^{(m_\Nex)}_{bc} 
\\[2pt]
\vdots & \vdots &  & \vdots & \vdots 
\\                                              
\up{wx}X^{(m_1)}_{li}   & \up{wx}X^{(m_2)}_{lj}  & \cdots& \up{ww}{X}^{(m_{\Nex-1})}_{ld} & \up{ww}{Y}^{(m_\Nex)}_{lc} 
\\[2pt]
\up{wx}X^{(m_1)}_{ki}   & \up{wx}X^{(m_2)}_{kj}  & \cdots& \up{ww}{X}^{(m_{\Nex-1})}_{kd} & \up{ww}{X}^{(m_\Nex)}_{kc} 
\end{vmatrix},
\end{split}
\label{eq:OVsplit}
\end{equation}
where $\Nex$ is the combined number of excitations in the bra and ket states.
Since the fundamental contractions $X$ and $Y$ can be precomputed and stored once for a given pair of 
reference determinants, the computational cost of evaluating Eq.~\eqref{eq:OVsplit} is only controlled by
the total number of excitations 
$\Nex$ and the number of zero-overlap orbitals in the biorthogonal basis $m$. 
Specifically, the cost of evaluating each determinant in Eq.~\eqref{eq:OVsplit} scales as $\Or{\Nex^3}$, while the number of 
ways to distribute the $m$ zeros over the $\Nex$ contractions is $\binom{\Nex}{m}$,
giving an overall scaling of $\Or{\Nex^3 \binom{\Nex}{m}}$.
In comparison, the conventional first-quantized approach evaluates the determinant of the 
occupied orbital overlap matrix between a pair of excited configurations, giving an $\Or{\Ne^3}$ scaling.\cite{MayerBook}
Since the number of excitations is smaller than the total number of electrons by definition,  
the evaluation of the extended nonorthogonal Wick's theorem using Eq.~\eqref{eq:OVsplit} is generally much more efficient.

\subsection{One-Body Coupling}
\label{subsec:one-body}
Representing one-body operators in the molecular orbital basis for determinant $\ket{\up{x}\Phi}$ as
\begin{equation}
\hat{f} = \sum_{pq} \up{x}f_{pq} \up{x}b^\dagger_p \up{x}b_q^{\vdag}
\end{equation}
allows a general coupling term between two excitations from nonorthogonal reference determinants to be expressed as
\begin{equation}
\begin{split}\mel*{ \up{x}\Phi_{ij\cdots}^{ab\dots} }{\hat{f}}{\up{w}\Phi_{kl\cdots}^{cd\dots} } 
=
\sum_{pq}
\,\up{x}f_{pq}
\bra{\up{x}\Phi} (\up{x}b_i^\dagger \up{x}b_a^{\vdag}) (\up{x}b_j^\dagger \up{x}b_b^{\vdag} )
\cdots 
(\up{x}b^\dagger_p \up{x}b_q^{\vdag})
\cdots
(\up{w}b_d^\dagger \up{w}b_l^{\vdag}) (\up{w}b_c^\dagger \up{w}b_k^{\vdag} )\ket{\up{w}\Phi} .
\end{split}
\end{equation}
Applying L\"{o}wdin's general formula for the extended nonorthogonal Wick's theorem give the expansion
\begin{equation}
\mel*{ \up{x}\Phi_{ij\cdots}^{ab\dots} }{\hat{f}}{\up{w}\Phi_{kl\cdots}^{cd\dots} } 
= \up{xw}\tS
 \hspace{-0.5em}
\sum_{\substack{m_1, \dots, m_{\Nex+1} \\ m_1 + \dots + m_{\Nex+1} = m}}
\sum_{pq}
\up{x}f_{pq}
\begin{vmatrix}
\up{xx}X_{ai}^{(m_1)}   & \up{xx}Y_{aj}^{(m_2)}  & \cdots & \up{xx}Y_{ap}^{(m_{\Nex_x+1})}  & \cdots &\up{xw}Y_{ad}^{(m_{\Nex})} & \up{xw}Y_{ac}^{(m_{\Nex+1})} \\
\up{xx}X_{bi}^{(m_1)}   & \up{xx}X_{bj}^{(m_2)} & \cdots & \up{xx}Y_{bp}^{(m_{\Nex_x+1})}   & \cdots &\up{xw}Y_{bd}^{(m_{\Nex})} & \up{xw}Y_{bc}^{(m_{\Nex+1})} \\
\vdots & \vdots &  & \vdots &  &  \vdots & \vdots \\
\up{xx}X_{qi}^{(m_1)} & \up{xx}X_{qj}^{(m_2)} & \cdots & \up{xx}X_{qp}^{(m_{\Nex_x+1})}    & \cdots & \up{xw}Y_{qd}^{(m_{\Nex})} & \up{xw}Y_{qc}^{(m_{\Nex+1})} \\
\vdots & \vdots &  & \vdots &  & \vdots & \vdots \\
\up{wx}X_{li}^{(m_1)}   & \up{wx}X_{lj}^{(m_2)}  & \cdots & \up{wx}X_{lp}^{(m_{\Nex_x+1})}  & \cdots &\up{ww}X_{ld}^{(m_{\Nex})} & \up{ww}Y_{lc}^{(m_{\Nex+1})} \\
\up{wx}X_{ki}^{(m_1)}  & \up{wx}X_{kj}^{(m_2)} & \cdots & \up{wx}X_{kp}^{(m_{\Nex_x+1})}   & \cdots &\up{ww}X_{kd}^{(m_{\Nex})} & \up{ww}X_{kc}^{(m_{\Nex+1})}
\end{vmatrix},
\label{eq:BigExpansion}
\end{equation}
where $\Nex_x$ is the number of excitations in the bra state and $\Nex$ is the combined 
number of excitations in the bra and ket states.
Following Ref.~\onlinecite{Lowdin1955a}, a Laplace expansion can then be performed along the row corresponding
to the one-body operator index $q$ to give
\begin{equation}
\begin{split}
\mel*{ \up{x}\Phi_{ij\cdots}^{ab\dots} }{\hat{f}}{\up{w}\Phi_{kl\cdots}^{cd\dots} } 
&= \up{xw}\tS
 \hspace{-0.5em}
\sum_{\substack{m_1, \dots, m_{\Nex+1} \\ m_1 + \dots + m_{\Nex+1} = m}}
\sum_{pq}
\up{x}f_{pq}\, \up{xx}X_{qp}^{(m_1)}
\begin{vmatrix}
\up{xx}X_{ai}^{(m_2)}   & \up{xx}Y_{aj}^{(m_3)}  & \cdots & \up{xw}Y_{ad}^{(m_{\Nex})} & \up{xw}Y_{ac}^{(m_{\Nex+1})} \\
\up{xx}X_{bi}^{(m_2)}   & \up{xx}X_{bj}^{(m_3)} & \cdots & \up{xw}Y_{bd}^{(m_{\Nex})} & \up{xw}Y_{bc}^{(m_{\Nex+1})} \\
\vdots & \vdots & & \vdots & \vdots \\
\up{wx}X_{li}^{(m_2)}   & \up{wx}X_{lj}^{(m_3)}  & \cdots & \up{ww}X_{ld}^{(m_{\Nex})} & \up{ww}Y_{lc}^{(m_{\Nex+1})} \\
\up{wx}X_{ki}^{(m_2)}  & \up{wx}X_{kj}^{(m_3)} & \cdots & \up{ww}X_{kd}^{(m_{\Nex})} & \up{ww}X_{kc}^{(m_{\Nex+1})}
\end{vmatrix}
\\[5pt]
&-\
\up{xw}\tS
\hspace{-0.5em}
\sum_{\substack{m_1, \dots, m_{\Nex+1} \\ m_1 + \dots + m_{\Nex+1} = m}}
\sum_{pq}
\up{xx}f_{pq} \up{xx}X_{qi}^{(m_1)} 
\begin{vmatrix}
\up{xx}Y_{ap}^{(m_2)} & \up{xx}Y_{aj}^{(m_3)}   &  \cdots   & \up{xw}Y_{ad}^{(m_\Nex)} & \up{xw}Y_{ac}^{(m_{\Nex+1})} \\
\up{xx}Y_{bp}^{(m_2)} & \up{xx}X_{bj}^{(m_3)}   &  \cdots   & \up{xw}Y_{bd}^{(m_\Nex)} & \up{xw}Y_{bc}^{(m_{\Nex+1})} \\
\vdots & \vdots & & \vdots & \vdots \\
\up{wx}X_{lp}^{(m_2)} & \up{wx}X_{lj}^{(m_3)}   & \cdots    & \up{ww}X_{ld}^{(m_\Nex)} & \up{ww}Y_{lc}^{(m_{\Nex+1})} \\
\up{wx}X_{kp}^{(m_2)} & \up{wx}X_{kj}^{(m_3)}   & \cdots    & \up{ww}X_{kd}^{(m_\Nex)} & \up{ww}X_{kc}^{(m_{\Nex+1})}
\end{vmatrix}
\\
&-
\cdots
\\
&-
\up{xw}\tS
\hspace{-0.5em}
\sum_{\substack{m_1, \dots, m_{\Nex+1} \\ m_1 + \dots + m_{\Nex+1} = m}}
\sum_{pq}
\up{x}f_{pq}\,  \up{xw}X_{qc}^{(m_1)}
\begin{vmatrix}
\up{xx}X_{ai}^{(m_2)} & \up{xx}Y_{aj}^{(m_3)}  & \cdots   & \up{xw}Y_{ad}^{(m_\Nex)} & \up{xx}Y_{ap}^{(m_{\Nex+1})} \\
\up{xx}X_{bi}^{(m_2)} & \up{xx}X_{bj}^{(m_3)}  & \cdots   & \up{xw}Y_{bd}^{(m_\Nex)} & \up{xx}Y_{bp}^{(m_{\Nex+1})} \\
\vdots                & \vdots                 &          & \vdots                & \vdots                \\
\up{wx}X_{li}^{(m_2)} & \up{wx}X_{lj}^{(m_3)}  & \cdots   & \up{ww}X_{ld}^{(m_\Nex)} & \up{wx}X_{lp}^{(m_{\Nex+1})} \\
\up{wx}X_{ki}^{(m_2)} & \up{wx}X_{kj}^{(m_3)}  & \cdots   & \up{ww}X_{kd}^{(m_\Nex)} & \up{wx}X_{kp}^{(m_{\Nex+1})}
\end{vmatrix},
\end{split}
\label{eq:onebody2}
\end{equation}
where, for convenience, the element  along this row corresponding to operator index $p$ 
is placed first and the dummy variables $m_k$ have been rearranged.
On the second row of Eq.~\eqref{eq:onebody2}, and those below it, 
the columns in the minor submatrices have been swapped so that the column 
corresponding to the one-body operator (index $p$) occupies the position originally held 
by the corresponding cofactor (\eg{}, index $i,j,d$ or $c$),
ensuring that these rows all have the same overall sign contribution.

While a na\"{i}ve implementation of Eq.~\eqref{eq:onebody2} scales as $\Or{\Nbas^2}$ due to the summations over $p$ and $q$,
this scaling can be removed by introducing a series of intermediate quantities. 
The first line requires the intermediate
\begin{align}
\up{xx}\cF_{0}^{(m_i)} &=\sum_{pq}   \up{x}f_{pq}\, \up{xx}X_{qp}^{(m_i)},
\end{align}
which takes different values depending on whether the contraction $\up{xx}X_{qp}^{(m_i)}$ is assigned to a zero-overlap orbital pair, 
as determined by $m_i$.
For the remaining lines in Eq.~\eqref{eq:onebody2}, the rules for multiplying a determinant by a scalar can be exploited to
move the summation over $p$ and $q$ inside the corresponding minor submatrices.
Introducing the intermediate quantities
\begin{subequations}
\begin{align}
\up{xx}\cF_{ab}^{(m_i,m_j)}  &= \sum_{pq}  \up{xx}Y_{ap}^{(m_i)}\,  \up{xx}f_{pq} \up{xx}X_{qb}^{(m_j)},
\\
\up{wx}\cF_{ab}^{(m_i,m_j)}  &= \sum_{pq}  \up{wx}X_{ap}^{(m_i)}\,  \up{xx}f_{pq} \up{xx}X_{qb}^{(m_j)},
\\
\up{xw}\cF_{ab}^{(m_i,m_j)}  &= \sum_{pq}  \up{xx}Y_{ap}^{(m_i)}\,  \up{xx}f_{pq} \up{xw}Y_{qb}^{(m_j)},
\\
\up{ww}\cF_{ab}^{(m_i,m_j)}  &= \sum_{pq}  \up{wx}X_{ap}^{(m_i)}\,  \up{xx}f_{pq} \up{xw}Y_{qb}^{(m_j)},
\end{align}
\label{eq:oneBodyInt}
\end{subequations}
and substituting these into the corresponding column for index $p$ gives the overall matrix element expression
\begin{equation}
\begin{split}
\mel*{ \up{x}\Phi_{ij\cdots}^{ab\dots} }{\hat{f}}{\up{w}\Phi_{kl\cdots}^{cd\dots} } 
&= \up{xw}\tS
 \hspace{-0.5em}
\sum_{\substack{m_1, \dots, m_{\Nex+1} \\ m_1 + \dots + m_{\Nex+1} = m}}
\up{xx}\cF_{0}^{(m_1)}
\begin{vmatrix}
\up{xx}X_{ai}^{(m_2)} & \up{xx}Y_{aj}^{(m_3)} & \cdots & \up{xw}Y_{ad}^{(m_{\Nex})} & \up{xw}Y_{ac}^{(m_{\Nex+1})} \\
\up{xx}X_{bi}^{(m_2)} & \up{xx}X_{bj}^{(m_3)} & \cdots & \up{xw}Y_{bd}^{(m_{\Nex})} & \up{xw}Y_{bc}^{(m_{\Nex+1})} \\
\vdots & \vdots & & \vdots & \vdots \\
\up{wx}X_{li}^{(m_2)} & \up{wx}X_{lj}^{(m_3)} & \cdots & \up{ww}X_{ld}^{(m_{\Nex})} & \up{ww}Y_{lc}^{(m_{\Nex+1})} \\
\up{wx}X_{ki}^{(m_2)} & \up{wx}X_{kj}^{(m_3)} & \cdots & \up{ww}X_{kd}^{(m_{\Nex})} & \up{ww}X_{kc}^{(m_{\Nex+1})}
\end{vmatrix}
\\
&-\
\up{xw}\tS
\hspace{-0.5em}
\sum_{\substack{m_1, \dots, m_{\Nex+1} \\ m_1 + \dots + m_{\Nex+1} = m}}
\
\begin{vmatrix}
\up{xx}\cF_{ai}^{(m_1,m_2)} & \up{xx}Y_{aj}^{(m_3)} &  \cdots &\up{xw}Y_{ad}^{(m_\Nex)} & \up{xw}Y_{ac}^{(m_{\Nex+1})} \\
\up{xx}\cF_{bi}^{(m_1,m_2)} & \up{xx}X_{bj}^{(m_3)} &  \cdots &\up{xw}Y_{bd}^{(m_\Nex)} & \up{xw}Y_{bc}^{(m_{\Nex+1})} \\
\vdots & \vdots & & \vdots & \vdots \\
\up{wx}\cF_{li}^{(m_1,m_2)} & \up{wx}X_{lj}^{(m_3)} & \cdots  &\up{ww}X_{ld}^{(m_\Nex)} & \up{ww}Y_{lc}^{(m_{\Nex+1})} \\
\up{wx}\cF_{ki}^{(m_1,m_2)} & \up{wx}X_{kj}^{(m_3)} & \cdots  &\up{ww}X_{kd}^{(m_\Nex)} & \up{ww}X_{kc}^{(m_{\Nex+1})}
\end{vmatrix}
\\
&-
\cdots
\\
&-
\up{xw}\tS
\hspace{-0.5em}
\sum_{\substack{m_1, \dots, m_{\Nex+1} \\ m_1 + \dots + m_{\Nex+1} = m}}
\
\begin{vmatrix}
\up{xx}X_{ai}^{(m_2)} & \up{xx}Y_{aj}^{(m_3)} & \cdots & \up{xw}Y_{ad}^{(m_{\Nex})} & \up{xw}\cF_{ac}^{(m_1,m_{\Nex+1})} \\
\up{xx}X_{bi}^{(m_2)} & \up{xx}X_{bj}^{(m_3)} & \cdots & \up{xw}Y_{bd}^{(m_{\Nex})} & \up{xw}\cF_{bc}^{(m_1,m_{\Nex+1})} \\
\vdots & \vdots & & \vdots & \vdots \\
\up{wx}X_{li}^{(m_2)} & \up{wx}X_{lj}^{(m_3)} & \cdots & \up{ww}X_{ld}^{(m_{\Nex})} & \up{ww}\cF_{lc}^{(m_1,m_{\Nex+1})}\\
\up{wx}X_{ki}^{(m_2)} & \up{wx}X_{kj}^{(m_3)} & \cdots & \up{ww}X_{kd}^{(m_{\Nex})} & \up{ww}\cF_{kc}^{(m_1,m_{\Nex+1})}
\end{vmatrix}.
\end{split}
\label{eq:onebody3}
\end{equation}
This expression comprises the term $\up{xx}\cF_0^{(m_1)}$ multiplied by the 
total overlap of the excited configurations, minus terms where each column of the submatrix
is replaced by the corresponding intermediate \eg{}, $\up{xw}\cF_{ai}^{(m_1,m_2)}$.

Computationally, the $\cF$ intermediates in Eq.~\eqref{eq:oneBodyInt} can be pre-computed with scaling 
$\Or{16n^3}$ and stored with scaling $\Or{16n^2}$, where the factor of 16 accounts
for the four possible values of $(m_i,m_j)$ and the four combinations $xx$, $xw$, $wx$, and $ww$.
Once these intermediates have been evaluated, the total cost of computing a
one-body matrix element only depends on the total number of 
excitations in the bra and ket states $\Nex$, which controls the size of the determinants in Eq.~\eqref{eq:onebody3}
and the number of columns that must be replaced with the intermediates, \eg{}\ $\up{xw}\cF_{ai}^{(m_1,m_2)}$. 
The cost of evaluating each determinant in Eq.~\eqref{eq:onebody3} scales as $\Or{\Nex^3}$ and there
are $\Nex+1$ determinants that must be computed.
The summation over the $m_i$ values depends on the total number of way 
to distribute $m$ zero-overlap orbitals between the $\Nex+1$ contractions, given by $\binom{\Nex+1}{m}$.
Consequently, the overall scaling for each one-body matrix element is
$\Or{\Nex^3 (\Nex+1) \binom{\Nex+1}{m}}$,
which is constant with respect to the number of electrons or the size of the basis set.
In comparison, applying the generalized Slater--Condon rules for each pair of determinants 
requires the biorthogonalisation of the \textit{full} set of occupied orbitals, with an 
iterative $\Or{\Ne^3}$ scaling, followed by an $\Or{n^2}$ contraction of the 
co-density matrices with the one-electron integrals.\cite{MayerBook,Thom2009}
Evidently, the new approach described here is significantly more efficient when a large
number of coupling elements are required.

\subsection{Two-Body Coupling}
\label{subsec:two-body}

Finally, two-body operators can be represented as
\begin{equation}
\hat{v} = \sum_{pqrs} \up{x}v_{prqs} (\up{x}b^\dagger_p \up{x}b_r^{\vdag}) (\up{x}b^\dagger_q \up{x}b_s^{\vdag}),
\end{equation}
where the two-electron integrals $\up{x}v_{prqs} = \,\up{x}(pr|qs)$ are expressed in the MO basis 
for reference $\ket{\up{x}\Phi}$ and are represented using Mulliken notation.\cite{SzaboBook}
A general coupling term between two excitations from nonorthogonal reference determinants is then given by
\begin{equation}
\begin{split}
\mel*{ \up{x}\Phi_{ij\cdots}^{ab\dots} }{\hat{v}}{\up{w}\Phi_{kl\cdots}^{cd\dots} } 
=
\sum_{pqrs}
 \up{x}v_{prqs}
\bra{\up{x}\Phi} (\up{x}b_i^\dagger \up{x}b_a^{\vdag}) (\up{x}b_j^\dagger \up{x}b_b^{\vdag} )
\cdots 
(\up{x}b^\dagger_p \up{x}b_r^{\vdag})
(\up{x}b^\dagger_q \up{x}b_s^{\vdag})
\cdots
(\up{w}b_d^\dagger \up{w}b_l^{\vdag}) (\up{w}b_c^\dagger \up{w}b_k^{\vdag} )\ket{\up{w}\Phi} .
\end{split}
\label{eq:twobody0}
\end{equation}
Deriving expressions for these matrix elements using the Laplace expansion approach is much more involved than 
the one-body operators and is much harder for the reader to follow.
Instead, an alternative approach 
can be used that constructs effective zero- or one-body operators by partially contracting subunits containing 
two or four of the indices $p,r,q,s$, 
and the one-body framework established in Section~\ref{subsec:one-body} can then be applied.
The final expression is given by 
\begin{equation}
\mel*{ \up{x}\Phi_{ij\cdots}^{ab\dots} }{\hat{v}}{\up{w}\Phi_{kl\cdots}^{cd\dots} } 
= 
\text{Eq.~\eqref{eq:twobody1} + Eq.~\eqref{eq:twobody2} + Eq.~\eqref{eq:twobody3}},
\end{equation}
where each constituent equation is derived in detail below.

First,  contractions containing the subunits 
$\wick{\up{x}\c1{b}_{p}^{\dagger} \up{x}\c1{b}_{r}^{\vdag} \up{x}\c1{b}_{q}^{\dagger} \up{x}\c1{b}_{s}^{\vdag}} $ and 
$\wick{\up{x}\c2{b}_{p}^{\dagger} \up{x}\c1{b}_{r}^{\vdag} \up{x}\c1{b}_{q}^{\dagger} \up{x}\c2{b}_{s}^{\vdag}} $ 
lead to the ``zeroth-order'' contribution
\begin{equation}
\up{xw}\tS
 \hspace{-0.5em}
\sum_{\substack{m_1, \dots, m_{\Nex+2} \\ m_1 + \dots + m_{\Nex+2} = m}}
\up{xx}\cV_0^{(m_1,m_2)}
\begin{vmatrix}
\up{xx}X_{ai}^{(m_3)}   & \up{xx}Y_{aj}^{(m_4)}  & \cdots & \up{xw}Y_{ad}^{(m_{\Nex+1})} & \up{xw}Y_{ac}^{(m_{\Nex+2})} \\
\up{xx}X_{bi}^{(m_3)}   & \up{xx}X_{bj}^{(m_4)} & \cdots & \up{xw}Y_{bd}^{(m_{\Nex+1})} & \up{xw}Y_{bc}^{(m_{\Nex+2})} \\
\vdots & \vdots & & \vdots & \vdots \\
\up{wx}X_{li}^{(m_3)}   & \up{wx}X_{lj}^{(m_4)}  & \cdots & \up{ww}X_{ld}^{(m_{\Nex+1})} & \up{ww}Y_{lc}^{(m_{\Nex+2})} \\
\up{wx}X_{ki}^{(m_3)}  & \up{wx}X_{kj}^{(m_4)} & \cdots & \up{ww}X_{kd}^{(m_{\Nex+1})} & \up{ww}X_{kc}^{(m_{\Nex+2})}
\end{vmatrix},
\label{eq:twobody1}
\end{equation}
where the intermediate terms have been defined
\begin{subequations}
\begin{align}
 \up{x}\tilde{v}_{pr}^{(m_i)} &=\sum_{qs} (\up{x}v_{prqs} -  \,\up{x}v_{psqr})\, \up{xx}X_{sq}^{(m_i)}
\label{eq:effectiveTwo}
\\
\up{xx}\cV_0^{(m_i,m_j)} &= \sum_{pr} \up{x}\tilde{v}_{pr}^{(m_i)}\, \up{xx}X_{rp}^{(m_j)}.
\end{align}
\end{subequations}
The contribution from Eq.~\eqref{eq:twobody1} is analogous to the one-body $\up{xx}\cF_0^{(m_1)}$ 
contribution in the first line in Eq.~\eqref{eq:onebody3}.

Next, the partially-contracted subunits 
$\wick{\up{x}\c1{b}_{p}^{\dagger} \up{x}\c1{b}_{r}^{\vdag} \up{x}{b}_{q}^{\dagger} \up{x}{b}_{s}^{\vdag}} $,
$\wick{\up{x}\c1{b}_{p}^{\dagger} \up{x}{b}_{r}^{\vdag} \up{x}{b}_{q}^{\dagger} \up{x}\c1{b}_{s}^{\vdag}} $,
$\wick{\up{x}{b}_{p}^{\dagger} \up{x}\c1{b}_{r}^{\vdag} \up{x}\c1{b}_{q}^{\dagger} \up{x}{b}_{s}^{\vdag}} $,
and
$\wick{\up{x}{b}_{p}^{\dagger} \up{x}{b}_{r}^{\vdag} \up{x}\c1{b}_{q}^{\dagger} \up{x}\c1{b}_{s}^{\vdag}} $,
can be used to define a new effective one-body operator
\begin{equation}
\hat{\tilde{v}}^{(m_i)} = 2 \sum_{pr} \ \up{x}\tilde{v}^{(m_i)}_{pr} \up{x}{b}_{p}^{\dagger} \up{x}{b}_{r}^{\vdag},
\end{equation}
where the intermediate term in Eq.~\eqref{eq:effectiveTwo} has been employed and 
the factor of two arises from the equivalence of terms such as 
$\wick{\up{x}\c1{b}_{p}^{\dagger} \up{x}\c1{b}_{r}^{\vdag} \up{x}{b}_{q}^{\dagger} \up{x}{b}_{s}^{\vdag}} $ and  
$\wick{\up{x}{b}_{p}^{\dagger} \up{x}{b}_{r}^{\vdag} \up{x}\c1{b}_{q}^{\dagger} \up{x}\c1{b}_{s}^{\vdag}} $
under the permutation of the dummy indices $p,q,r,s$.
The contribution from these partially contracted subunits can then be evaluated using the rules established 
for one-body operators in Section~\ref{subsec:one-body} by introducing the additional intermediate terms
\begin{subequations}
\begin{align}
\up{xx}\cV_{ab}^{(m_1,m_2,m_3)}  &= \sum_{pq}  \up{xx}Y_{ap}^{(m_1)}\,  \up{x}\tilde{v}^{(m_2)}_{pq}\, \up{xx}X_{qb}^{(m_3)},
\\
\up{wx}\cV_{ab}^{(m_1,m_2,m_3)}  &= \sum_{pq}  \up{wx}X_{ap}^{(m_1)}\, \up{x}\tilde{v}^{(m_2)}_{pq}\, \up{xx}X_{qb}^{(m_3)},
\\
\up{xw}\cV_{ab}^{(m_1,m_2,m_3)}  &= \sum_{pq}  \up{xx}Y_{ap}^{(m_1)}\,  \up{x}\tilde{v}^{(m_2)}_{pq}\, \up{xw}Y_{qb}^{(m_3)},
\\
\up{ww}\cV_{ab}^{(m_1,m_2,m_3)}  &= \sum_{pq}  \up{wx}X_{ap}^{(m_1)}\, \up{x}\tilde{v}^{(m_2)}_{pq}\, \up{xw}Y_{qb}^{(m_3)}.
\end{align}
\label{eq:twoBodyInt2}
\end{subequations}
Each of these intermediates can be computed with a computational scaling of $\Or{8\Nbas^3}$ and 
the overall storage cost for a pair of reference determinants is $32\Nbas^2$.
The one-body terms that are analogous to the first line in Eq.~\eqref{eq:onebody3}, which represent the contractions 
$\wick{\up{x}\c1{b}_{p}^{\dagger} \up{x}\c1{b}_{r}^{\vdag} \up{x}\c1{b}_{q}^{\dagger} \up{x}\c1{b}_{s}^{\vdag}} $ and 
$\wick{\up{x}\c2{b}_{p}^{\dagger} \up{x}\c1{b}_{r}^{\vdag} \up{x}\c1{b}_{q}^{\dagger} \up{x}\c2{b}_{s}^{\vdag}} $,  
have already been included in Eq.~\eqref{eq:twobody1}.
Therefore, the unique contributions of these effective one-body operators 
to the full two-body matrix element in Eq.~\eqref{eq:twobody0} is
\begin{equation}
\begin{split}
-2\,\up{xw}\tS
\hspace{-0.5em}
\sum_{\substack{m_1, \dots, m_{\Nex+2} \\ m_1 + \dots + m_{\Nex+2} = m}}
\left(
\
\begin{vmatrix}
\up{xx}\cV_{ai}^{(m_1,m_2,m_3)} & \up{xx}Y_{aj}^{(m_4)} & \cdots & \up{xw}Y_{ad}^{(m_{\Nex+1})} & \up{xw}Y_{ac}^{(m_{\Nex+2})} \\
\up{xx}\cV_{bi}^{(m_1,m_2,m_3)} & \up{xx}X_{bj}^{(m_4)} & \cdots & \up{xw}Y_{bd}^{(m_{\Nex+1})} & \up{xw}Y_{bc}^{(m_{\Nex+2})} \\
\vdots & \vdots & & \vdots & \vdots \\
\up{wx}\cV_{li}^{(m_1,m_2,m_3)} & \up{wx}X_{lj}^{(m_4)} & \cdots & \up{ww}X_{ld}^{(m_{\Nex+1})} & \up{ww}Y_{lc}^{(m_{\Nex+2})} \\
\up{wx}\cV_{ki}^{(m_1,m_2,m_3)} & \up{wx}X_{kj}^{(m_4)} & \cdots & \up{ww}X_{kd}^{(m_{\Nex+1})} & \up{ww}X_{kc}^{(m_{\Nex+2})}
\end{vmatrix}
+\cdots+
\begin{vmatrix}
\up{xx}X_{ai}^{(m_1)} & \up{xx}Y_{aj}^{(m_2)} & \cdots & \up{xw}Y_{ad}^{(m_{\Nex-1})} & \up{xw}\cV_{ac}^{(m_\Nex,m_{\Nex+1},m_{\Nex+2})} \\
\up{xx}X_{bi}^{(m_1)} & \up{xx}X_{bj}^{(m_2)} & \cdots & \up{xw}Y_{bd}^{(m_{\Nex-1})} & \up{xw}\cV_{bc}^{(m_\Nex,m_{\Nex+1},m_{\Nex+2})} \\
\vdots & \vdots & & \vdots & \vdots \\
\up{wx}X_{li}^{(m_1)} & \up{wx}X_{lj}^{(m_2)} & \cdots & \up{ww}X_{ld}^{(m_{\Nex-1})} & \up{ww}\cV_{lc}^{(m_\Nex,m_{\Nex+1},m_{\Nex+2})}\\
\up{wx}X_{ki}^{(m_1)} & \up{wx}X_{kj}^{(m_2)} & \cdots & \up{ww}X_{kd}^{(m_{\Nex-1})} & \up{ww}\cV_{kc}^{(m_\Nex,m_{\Nex+1},m_{\Nex+2})}
\end{vmatrix}
\right).
\end{split}
\label{eq:twobody2}
\end{equation}
The summation within the parentheses includes the replacement of each column of the overlap determinant
with the corresponding intermediate terms from Eq.~\eqref{eq:twoBodyInt2}.

The remaining terms include cases where all the indices $p,q,r,s$ are contracted with operators 
that occur in the bra or ket excitation strings.
An effective one-body operator can be constructed by considering partial contractions with the form \eg{},
\begin{equation}
\begin{split}
&\bra{\up{x}\Phi}\wick{
(\up{x}{b}_{i}^{\dagger} \up{x}\c1{b}_{a}^{\vdag})
(\up{x}{b}_{j}^{\dagger} \up{x}{b}_{b}^{\vdag})
\cdots
(\up{x}\c1{b}_{p}^{\dagger} \up{x}\c1{b}_{r}^{\vdag}) 
(\up{x}{b}_{q}^{\dagger} \up{x}{b}_{s}^{\vdag})
\cdots 
(\up{w}\c1{b}_{d}^{\dagger} \up{w}{b}_{l}^{\vdag})
(\up{w}{b}_{c}^{\dagger} \up{w}{b}_{k}^{\vdag}) 
}
\ket{\up{w}\Phi}
\\
+
&\bra{\up{x}\Phi}\wick{
(\up{x}{b}_{i}^{\dagger} \up{x}\c1{b}_{a}^{\vdag})
(\up{x}{b}_{j}^{\dagger} \up{x}{b}_{b}^{\vdag})
\cdots
(\up{x}\c1{b}_{p}^{\dagger} \up{x}{b}_{r}^{\vdag}) 
(\up{x}{b}_{q}^{\dagger} \up{x}\c1{b}_{s}^{\vdag})
\cdots 
(\up{w}\c1{b}_{d}^{\dagger} \up{w}{b}_{l}^{\vdag})
(\up{w}{b}_{c}^{\dagger} \up{w}{b}_{k}^{\vdag}) 
}
\ket{\up{w}\Phi},
\end{split}
\end{equation}
where the indices $p,r$ and $p,s$ are contracted with bra or ket excitation operators,
giving effective operators with the form \eg{},
\begin{subequations}
\begin{align}
\up{xx}\hat{J}_{ab}^{\,(m_i,m_j)} 
&= \phi_{ab} \sum_{sq} 
\up{xx}Y_{ap}^{(m_i)}\, 
(\up{x}v_{prqs} - \up{x}v_{psqr})\, 
\up{xx}X_{rb}^{(m_j)}\,
\up{x}b^\dagger_q \up{x}b_s^{\vdag},
\\                                    
\up{wx}\hat{J}_{ab}^{\,(m_i,m_j)} 
&= \phi_{ab} \sum_{sq} 
\up{wx}X_{ap}^{(m_i)}\, 
(\up{x}v_{prqs} - \up{x}v_{psqr})\, 
\up{xx}X_{rb}^{(m_j)}\,
\up{x}b^\dagger_q \up{x}b_s^{\vdag},
\\                                   
\up{xw}\hat{J}_{ab}^{\,(m_i,m_j)} 
&= \phi_{ab} \sum_{sq} 
\up{xx}Y_{ap}^{(m_i)}\, 
(\up{x}v_{prqs} - \up{x}v_{psqr})\, 
\up{xw}Y_{rb}^{(m_j)}\,
\up{x}b^\dagger_q \up{x}b_s^{\vdag},
\\                                    
\up{ww}\hat{J}_{ab}^{\,(m_i,m_j)} 
&= \phi_{ab} \sum_{sq} 
\up{wx}X_{ap}^{(m_i)}\, 
(\up{x}v_{prqs} - \up{x}v_{psqr})\, 
\up{xw}Y_{rb}^{(m_j)}\,
\up{x}b^\dagger_q \up{x}b_s^{\vdag},
\end{align}
\label{eq:effectivetwo}
\end{subequations}
where the permutation of the dummy indices $p,q,r,s$ has been exploited 
to incorporate the ``exchange-like'' terms.
The operators in Eq.~\eqref{eq:effectivetwo} contain a phase factor $\phi_{ab}$ that 
takes the value  $+1$ if $a$ and $b$ correspond to the same excitation or 
are separated by an odd number of excitations, and -1 if $a$ and $b$ 
correspond to different excitations separated by an even number of excitations,  
as illustrated for the example 
$\mel*{ \up{x}\Phi_{ij}^{ab} }{\hat{v}}{\up{w}\Phi_{kl}^{cd} } $ by the chequerboard pattern
\begin{equation*}
\phi = \left\lbrace\,
\begin{matrix}
 & i & j &  d & c
\\ 
a & + & - &  + & -
\\
b & -  & + & - & +
\\
l & + & - & + & -
\\
k & - & + & - & +
\end{matrix}
\right.
.
\end{equation*}
Every effective one-body operator defined for the $\Nex^2$ combinations of a creation and annihilation 
operator pair (excluding $p,q,r,s$) must be considered. 
The contribution to the full two-body matrix element can then be evaluated using the one-body approach 
described in Section~\ref{subsec:one-body} by introducing a final set of intermediates
\begin{subequations}
\begin{align}
\up{xx,xx}\cJ_{ab,cd}^{(m_i,m_j,m_k,m_l)} 
&= \sum_{pqrs}(\up{x}v_{prqs}- \up{x}v_{psqr}) \, 
\up{xx}Y_{ap}^{(m_i)}\, \up{xx}X_{rb}^{(m_j)}\, \up{xx}Y_{cq}^{(m_k)}\, \up{xx}X_{sd}^{(m_l)}
\\
\up{xx,xw}\cJ_{ab,cd}^{(m_i,m_j,m_k,m_l)} 
&= \sum_{pqrs}(\up{x}v_{prqs}- \up{x}v_{psqr}) \, 
\up{xx}Y_{ap}^{(m_i)}\, \up{xx}X_{rb}^{(m_j)}\, \up{xx}Y_{cq}^{(m_k)}\, \up{xw}Y_{sd}^{(m_l)}
\\
\up{xx,wx}\cJ_{ab,cd}^{(m_i,m_j,m_k,m_l)} 
&= \sum_{pqrs}(\up{x}v_{prqs}- \up{x}v_{psqr}) \, 
\up{xx}Y_{ap}^{(m_i)}\, \up{xx}X_{rb}^{(m_j)}\, \up{wx}X_{cq}^{(m_k)}\, \up{xx}X_{sd}^{(m_l)}
\\
\up{xx,ww}\cJ_{ab,cd}^{(m_i,m_j,m_k,m_l)} 
&= \sum_{pqrs}(\up{x}v_{prqs}- \up{x}v_{psqr}) \, 
\up{xx}Y_{ap}^{(m_i)}\, \up{xx}X_{rb}^{(m_j)}\, \up{wx}X_{cq}^{(m_k)}\, \up{xw}Y_{sd}^{(m_l)}
\\
\up{xw,xw}\cJ_{ab,cd}^{(m_i,m_j,m_k,m_l)} 
&= \sum_{pqrs}(\up{x}v_{prqs}- \up{x}v_{psqr}) \, 
\up{xx}Y_{ap}^{(m_i)}\, \up{xw}Y_{rb}^{(m_j)}\, \up{xx}Y_{cq}^{(m_k)}\, \up{xw}Y_{sd}^{(m_l)}
\\
\up{xw,wx}\cJ_{ab,cd}^{(m_i,m_j,m_k,m_l)} 
&= \sum_{pqrs}(\up{x}v_{prqs}- \up{x}v_{psqr}) \, 
\up{xx}Y_{ap}^{(m_i)}\, \up{xw}Y_{rb}^{(m_j)}\, \up{wx}X_{cq}^{(m_k)}\, \up{xx}X_{sd}^{(m_l)}
\\
\up{wx,wx}\cJ_{ab,cd}^{(m_i,m_j,m_k,m_l)} 
&= \sum_{pqrs}(\up{x}v_{prqs}- \up{x}v_{psqr}) \, 
\up{wx}X_{ap}^{(m_i)}\, \up{xx}X_{rb}^{(m_j)}\, \up{wx}X_{cq}^{(m_k)}\, \up{xx}X_{sd}^{(m_l)}
\\
\up{ww,wx}\cJ_{ab,cd}^{(m_i,m_j,m_k,m_l)} 
&= \sum_{pqrs}(\up{x}v_{prqs}- \up{x}v_{psqr}) \, 
\up{wx}X_{ap}^{(m_i)}\, \up{xw}Y_{rb}^{(m_j)}\, \up{wx}X_{cq}^{(m_k)}\, \up{xx}X_{sd}^{(m_l)}
\\
\up{ww,xw}\cJ_{ab,cd}^{(m_i,m_j,m_k,m_l)} 
&= \sum_{pqrs}(\up{x}v_{prqs}- \up{x}v_{psqr}) \, 
\up{wx}X_{ap}^{(m_i)}\, \up{xw}Y_{rb}^{(m_j)}\, \up{xx}Y_{cq}^{(m_k)}\, \up{xw}Y_{sd}^{(m_l)}
\\
\up{ww,ww}\cJ_{ab,cd}^{(m_i,m_j,m_k,m_l)} 
&= \sum_{pqrs}(\up{x}v_{prqs}- \up{x}v_{psqr}) \, 
\up{wx}X_{ap}^{(m_i)}\, \up{xw}Y_{rb}^{(m_j)}\, \up{wx}X_{cq}^{(m_k)}\, \up{xw}Y_{sd}^{(m_l)}.
\end{align}
\label{eq:TwoBodyInt2}
\end{subequations}
The standard two-electron integral symmetry relation
$\up{xw,yz}\cJ_{ab,cd}^{(m_i,m_j,m_k,m_l)} = \up{yz,xw}\cJ_{cd,ab}^{(m_k,m_l,m_i,m_j)}$ 
allows intermediates for the remaining combinations of indices to be obtained.
Each intermediate can be computed with a computational scaling of $\Or{16\Nbas^5}$,
where the factor of 16 comes from the possible combinations of $(m_i,m_j,m_k,m_l)$ 
and the maximum storage requirement is $256\Nbas^4$.
In practice, this storage requirement is reduced if not all combinations 
of $(m_i,m_j,m_k,m_l)$ are required (\ie{}, $m<4$) 
or if excitations are only considered within an active orbital space.

When applying the one-body framework (Section~\ref{subsec:one-body}) to the effective operators 
defined in Eq.~\eqref{eq:effectivetwo},
the terms that arise from contracting the $q$ and $s$ indices are discarded 
as these are already 
taken into account by Eq.~\eqref{eq:twobody2}.
These terms correspond to the first line in Eq.~\eqref{eq:onebody3}. 
Therefore, the remaining unique contributions to the two-body matrix element are 
\begin{equation}
\begin{split}
&-\frac{\phi_{ai}}{2}\,\up{xw}\tS
\hspace{-0.5em}
\sum_{\substack{m_1, \dots, m_{\Nex+2} \\ m_1 + \dots + m_{\Nex+2} = m}}
\left(
\
\begin{vmatrix}
\up{xx,xx}\cJ_{ai,bj}^{(m_1,m_2,m_3,m_4)}  & \cdots &\up{xw}Y_{bd}^{(m_{\Nex+1})} & \up{xw}Y_{bc}^{(m_{\Nex+2})} \\
\vdots                                     &        &\vdots                    & \vdots                      \\
\up{xx,wx}\cJ_{ai,lj}^{(m_1,m_2,m_3,m_4)}  & \cdots &\up{ww}X_{ld}^{(m_{\Nex+1})} & \up{ww}Y_{lc}^{(m_{\Nex+2})} \\
\up{xx,wx}\cJ_{ai,kj}^{(m_1,m_2,m_3,m_4)}  & \cdots &\up{ww}X_{kd}^{(m_{\Nex+1})} & \up{ww}X_{kc}^{(m_{\Nex+2})}
\end{vmatrix}
+\cdots+
\begin{vmatrix}
\up{xx}X_{bj}^{(m_1)} & \cdots &\up{xw}Y_{bd}^{(m_{\Nex-2})} & \up{xx,xw}\cJ_{ai,bc}^{(m_{\Nex-1},m_\Nex,m_{\Nex+1},m_{\Nex+2})} \\
\vdots                &        &\vdots                    & \vdots                                                \\
\up{wx}X_{lj}^{(m_1)} & \cdots &\up{ww}X_{ld}^{(m_{\Nex-2})} & \up{xx,ww}\cJ_{ai,lc}^{(m_{\Nex-1},m_\Nex,m_{\Nex+1},m_{\Nex+2})} \\
\up{wx}X_{kj}^{(m_1)} & \cdots &\up{ww}X_{kd}^{(m_{\Nex-2})} & \up{xx,ww}\cJ_{ai,kc}^{(m_{\Nex-1},m_\Nex,m_{\Nex+1},m_{\Nex+2})}
\end{vmatrix}
\right)
\\
&-\frac{\phi_{aj}}{2}\,\up{xw}\tS
\hspace{-0.5em}
\sum_{\substack{m_1, \dots, m_{\Nex+2} \\ m_1 + \dots + m_{\Nex+2} = m}}
\left(
\
\begin{vmatrix}
\up{xx,xx}\cJ_{aj,bi}^{(m_1,m_2,m_3,m_4)}  &  \cdots & \up{xw}Y_{bd}^{(m_{\Nex+1})} & \up{xw}Y_{bc}^{(m_{\Nex+2})} \\
\vdots                                     &         & \vdots                    & \vdots                    \\
\up{xx,wx}\cJ_{aj,li}^{(m_1,m_2,m_3,m_4)}  & \cdots  & \up{ww}X_{ld}^{(m_{\Nex+1})} & \up{ww}Y_{lc}^{(m_{\Nex+2})} \\
\up{xx,wx}\cJ_{aj,ki}^{(m_1,m_2,m_3,m_4)}  & \cdots  & \up{ww}X_{kd}^{(m_{\Nex+1})} & \up{ww}X_{kc}^{(m_{\Nex+2})}
\end{vmatrix}
+\cdots+
\begin{vmatrix}
\up{xx}X_{bi}^{(m_1)} & \cdots & \up{xw}Y_{bd}^{(m_{\Nex-2})} & \up{xx,xw}\cJ_{aj,bc}^{(m_{\Nex-1},m_\Nex,m_{\Nex+1},m_{\Nex+2})} \\
\vdots                &        & \vdots                    & \vdots                                                \\
\up{wx}X_{li}^{(m_1)} & \cdots & \up{ww}X_{ld}^{(m_{\Nex-2})} & \up{xx,ww}\cJ_{aj,lc}^{(m_{\Nex-1},m_\Nex,m_{\Nex+1},m_{\Nex+2})} \\
\up{wx}X_{ki}^{(m_1)} & \cdots & \up{ww}X_{kd}^{(m_{\Nex-2})} & \up{xx,ww}\cJ_{aj,kc}^{(m_{\Nex-1},m_\Nex,m_{\Nex+1},m_{\Nex+2})}
\end{vmatrix}
\right)
\\
&- \cdots
\\
&-\frac{\phi_{kc}}{2}\,\up{xw}\tS
\hspace{-0.5em}
\sum_{\substack{m_1, \dots, m_{\Nex+2} \\ m_1 + \dots + m_{\Nex+2} = m}}
\left(
\
\begin{vmatrix}
\up{ww,xx}\cJ_{kc,ai}^{(m_1,m_2,m_3,m_4)} & \up{xx}Y_{aj}^{(m_5)} & \cdots & \up{xw}Y_{ad}^{(m_{\Nex+2})} \\
\up{ww,xx}\cJ_{kc,bi}^{(m_1,m_2,m_3,m_4)} & \up{xx}X_{bj}^{(m_5)} & \cdots & \up{xw}Y_{bd}^{(m_{\Nex+2})} \\
\vdots                                    & \vdots                &        & \vdots                       \\
\up{ww,wx}\cJ_{kc,li}^{(m_1,m_2,m_3,m_4)} & \up{wx}X_{lj}^{(m_5)} & \cdots & \up{ww}X_{ld}^{(m_{\Nex+2})} \\
\end{vmatrix}
+\cdots+
\begin{vmatrix}
\up{xx}X_{ai}^{(m_1)} & \up{xx}Y_{aj}^{(m_2)} & \cdots & \up{ww,xw}\cJ_{kc,ad}^{(m_{\Nex-1},m_\Nex,m_{\Nex+1},m_{\Nex+2})} \\
\up{xx}X_{bi}^{(m_1)} & \up{xx}X_{bj}^{(m_2)} & \cdots & \up{ww,xw}\cJ_{kc,bd}^{(m_{\Nex-1},m_\Nex,m_{\Nex+1},m_{\Nex+2})} \\
\vdots                & \vdots                &        & \vdots                                                \\
\up{wx}X_{li}^{(m_1)} & \up{wx}X_{lj}^{(m_2)} & \cdots & \up{ww,ww}\cJ_{kc,ld}^{(m_{\Nex-1},m_\Nex,m_{\Nex+1},m_{\Nex+2})} \\
\end{vmatrix}
\right).
\end{split}
\label{eq:twobody3}
\end{equation}
Individual lines in this expression correspond to the effective one-body operators 
constructed from a different pair of
the creation and annihilation operators selected from excitation strings (i.e., $(ai), (aj), \dots, (kc)$).
The sum within a line corresponds to the replacement of each column by the corresponding intermediate 
for the remaining excitation indices, where the excitation indices used to build the effective 
one-body operators have been removed.
Every contribution also includes a summation over the possible ways to distribute the
$m$ zero-overlap orbital pairs over the contractions.
The factor of $1/2$ accounts for the double counting of contributions, for example the term 
$\up{xx,xx}\cJ_{ai,bj}^{(m_1,m_2,m_3,m_4)}$ appears for the effective one-body operator with 
indices $(ai)$ and $(bj)$ due to the symmetry 
$\up{xx,xx}\cJ_{ai,bj}^{(m_1,m_2,m_3,m_4)} =         \up{xx,xx}\cJ_{bj,ai}^{(m_1,m_2,m_3,m_4)}$.

Once all the required intermediates have been computed and stored, the computational cost of evaluating 
a two-body matrix element using this approach is determined the the cost of computing
Eqs.~\eqref{eq:twobody1}, \eqref{eq:twobody2} and \eqref{eq:twobody3}.
For each term, there are $\binom{\Nex+2}{m}$ ways to distribute $m$ zero-overlap orbital pairs among 
the $\Nex+2$ contractions.
Eq.~\eqref{eq:twobody1} involves the computation of only one $\Nex \times \Nex$ determinant, giving an
$\Or{\Nex^3\binom{\Nex+2}{m}}$ scaling.
Eq.~\eqref{eq:twobody2} requires the computation of $\Nex$ determinants
where each column in the overlap determinant is replaced by intermediates of the form Eq.~\eqref{eq:twoBodyInt2},
giving a scaling of $\Or{\Nex^4 \binom{\Nex+2}{m}}$. 
Finally, Eq.~\eqref{eq:twobody3} involves the computation of $\Nex-1$ determinants with dimensions 
$(\Nex-1)\times(\Nex-1)$ for each pair of creation and annihilation operators in the excitation strings, 
giving a scaling of $\Or{\Nex^2(\Nex-1)^4\binom{\Nex+2}{m}}$.
The asymptotic scaling for individual two-body matrix elements using these intermediates is therefore 
$\Or{\Nex^6\binom{\Nex+2}{m}}$.
Although this computational cost increases rapidly with the number of excitations in a 
two-body matrix element, it has $\Or{1}$ scaling with respect to the number of basis functions or electrons.
In contrast, applying the generalized Slater--Condon rules for two-body operators scales as $\Or{\Nbas^4}$ 
for each matrix element and rapidly makes the computation of a large number of matrix elements unfeasible. 
\end{widetext}

\section{Illustration of Computational Scaling}
\label{sec:results}

The extended nonorthogonal Wick's theorem for arbitrary excitations has been implemented in a developmental
open-source C++ package \LibGNME{}, available for download at Ref.~\onlinecite{LibGNME}.
The primary advantage of the extended nonorthogonal Wick's theorem over the generalized Slater--Condon rules
is the $\Or{1}$ scaling with respect to the number of basis functions or electrons 
that can be achieved once all the required intermediates have been pre-computed.
This scaling means that the computation of nonorthogonal matrix elements for excited configurations becomes
almost as straightforward as the conventional orthogonal Slater--Condon rules or Wick's theorem.

The acceleration relative to the generalized Slater--Condon rules can be demonstrated by
comparing the average time required to compute one- and two-body matrix elements between singly- or doubly-excited
configurations with increasingly large correlation-consistent basis sets.\cite{Dunning1989}
Illustrative calculations using \LibGNME{} have been performed for two nonorthogonal reference determinants
corresponding to the spin-flip pair of spin-symmetry-broken unrestricted Hartree--Fock 
solutions for the ground state of \ce{H2O} at a bond angle of 104.5 and $R(\ce{O-H}) = \SI{1.35}{\angstrom}$.

\begin{figure}[b]
\includegraphics[width=\linewidth]{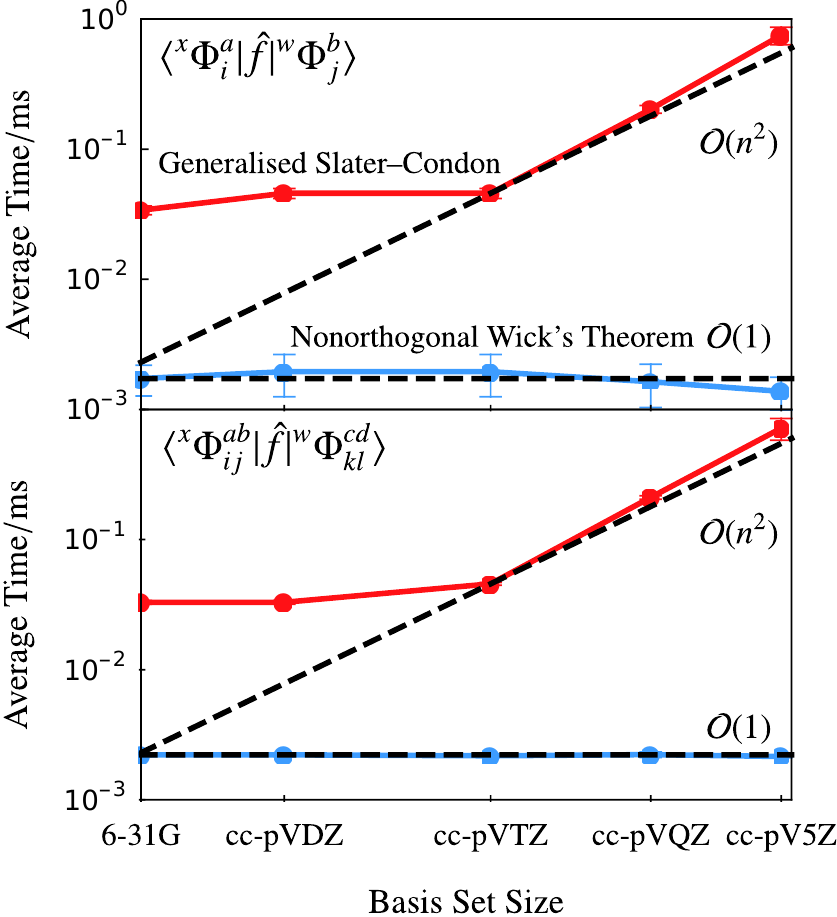}
\caption{Comparison of the average timing for computing one-body matrix elements in \ce{H2O}
using the extended nonorthogonal Wick's theorem (blue) and the generalized Slater--Condon rules (red).
Data are plotted on a log-log scale.}
\label{fig:onebody}
\end{figure}

Average timings for each one-body matrix element are compared between the extended nonorthogonal Wick's theorem
and the generalized Slater--Condon rules in Fig.~\ref{fig:onebody}.
This average is computed using all the singly- or doubly-excited coupling terms 
within a (13,10) active space comprising the lowest-energy molecular orbitals. 
The statistical distribution is assessed using 48 replicas of each calculation.
As expected, these data demonstrate the $\Or{1}$ scaling of the extended nonorthogonal Wick's theorem with respect 
to the basis set size, while the generalized Slater--Condon rules scale asymptotically as $\Or{\Nbas^2}$.
For small basis sets, the scaling of the generalized Slater--Condon rules 
becomes constant as the computational cost is dominated by biorthogonalising the occupied 
orbitals in each pair of excited configurations.
In the largest basis set considered (cc-pV5Z), the extended nonorthogonal Wick's theorem provides a computational 
acceleration of nearly 3 orders of magnitude relative to the generalized Slater--Condon rules.

\begin{figure}[tb]
\includegraphics[width=\linewidth]{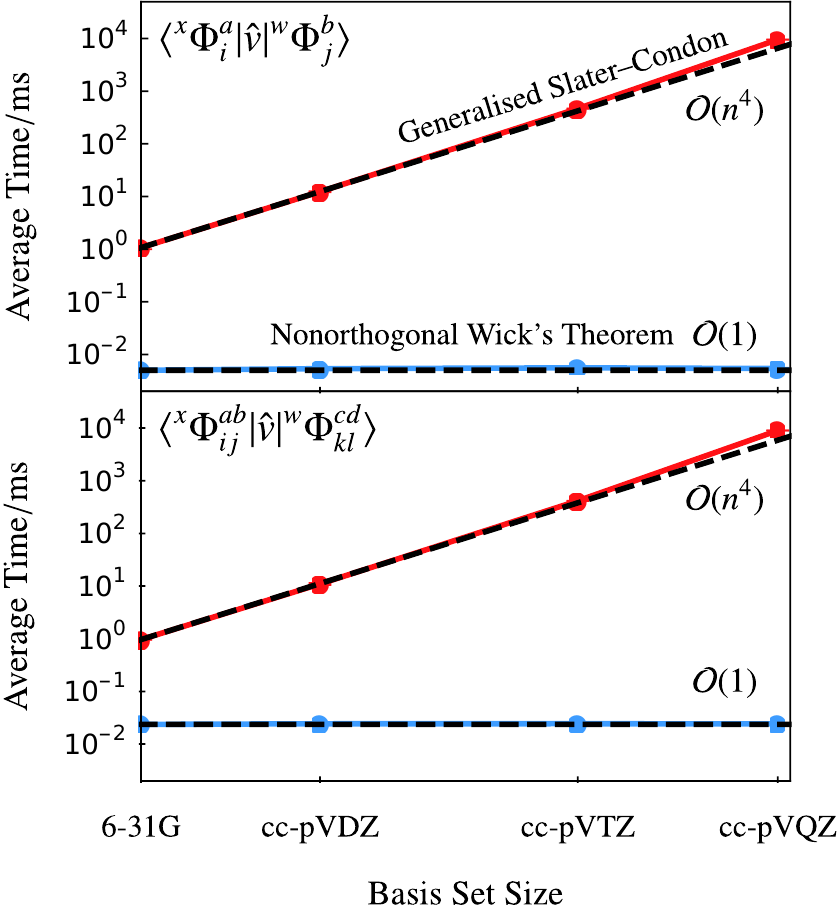}
\caption{Comparison of the average timing for computing two-body matrix elements in \ce{H2O} 
using the extended nonorthogonal Wick's theorem (blue) and the generalized Slater--Condon rules (red). 
Data are plotted on a log-log scale.}
\label{fig:twobody}
\end{figure}

Analogous average timings for two-body matrix elements between singly- and doubly-excited configurations are 
presented in Fig.~\ref{fig:twobody}.
Here, the generalized Slater--Condon rules show a near-ideal $\Or{\Nbas^4}$ computational scaling, while the 
extended nonorthogonal Wick's theorem has a constant cost for all basis sets, as predicted.
The larger $\Nbas^4$ scaling of two-body compared to one-body matrix elements means that the extended 
nonorthogonal Wick's theorem offers an even greater advantage over the generalized Slater--Condon rules, 
as demonstrated by the  six orders of magnitude acceleration 
achieved for the $\mel*{\up{x}\Phi_i^a}{\hat{b}}{\up{w}\Phi_{j}^{b}}$ matrix elements in the cc-pVQZ basis set.
These numerical results highlight the significant computational advance provided by the generalized nonorthogonal 
matrix elements compared to the previous state-of-the-art. 

\section{Discussion and outlook}
\label{sec:conc}

This work has extended the generalized nonorthogonal matrix elements introduced in 
Ref.~\onlinecite{Burton2021a} to derive coupling elements for arbitrary excited configurations.
By pre-computing and storing various intermediate terms, subsequent one- and two-body matrix elements 
can be evaluated with a computational cost that scales $\Or{1}$ with respect to the number of basis 
functions or electrons.
These developments provide a significant improvement over the commonly used generalized 
Slater--Condon rules, which asymptotically scale as $\Or{\Nbas^2}$ or $\Or{\Nbas^4}$ for
one- or two-body matrix elements, respectively.

While explicit expressions for certain nonorthogonal matrix elements have previously been reported,
this work presents an entirely general and unified framework for developing nonorthogonal techniques.
The current approach shares many similarities with the nonorthogonal matrix elements derived by 
Mahajan and Sharma,\cite{Mahajan2020,Mahajan2021} 
including the use of the determinantal expansion of Wick's theorem.
However, in contrast to their work, 
the derivations presented here are entirely generalized to cases where zero-overlap orbital pairs
occur in the biorthogonalisation of the reference determinants, providing a unification with the various
instances of the generalized Slater--Condon rules.\cite{MayerBook}
Consequently, this framework establishes the most general and flexible version of 
Wick's theorem for computing matrix elements between arbitrary Slater determinants.

In practice, the bottleneck for these generalized nonorthogonal matrix elements is the computation
and storage of the intermediate terms required for two-body operators.
While the evaluation of the intermediates in Eq.~\eqref{eq:TwoBodyInt2} has the same $\Or{\Nbas^5}$ 
scaling as a standard two-electron integral transformation, the cost of storing all these 
intermediates for a given pair of determinants scales as $256\, \Nbas^4$. 
The storage of two-electron integrals is already challenging for larger basis sets, so increasing this 
by a factor of 256 represents a significant computational overhead.
Fortunately, this cost can be reduced if there are no zero-overlap orbital pairs in the reference 
determinants (\ie{}, $m=0$) 
such that only a subset of $(m_i,m_j,m_k,m_l)$ combinations are required, or if excitations are only considered 
within an active orbital space.
Alternatively, techniques such as the Cholesky decomposition\cite{Koch2003} or 
resolution-of-the-identity\cite{Kendall1997} may provide further computational savings.

Until now, the development of advanced  nonorthogonal techniques in electronic structure 
theory has been hindered by the computational cost of computing arbitrary nonorthogonal matrix elements. 
The generalized framework introduced in Ref.~\onlinecite{Burton2021a}, and extended in this work, 
now offers a practical route to overcome this challenge. 
Moving forwards, the ability to rapidly compute nonorthogonal matrix elements between excited 
configurations will allow on-the-fly implementations of NOCI extensions such as NOCI-PT2,\cite{Burton2020} 
NOCI-MP2,\cite{Yost2016,Yost2019} and NOCIS.\cite{Oosterbaan2018,Oosterbaan2019}
Furthermore, the generalisation to arbitrary excitations will enable the 
evaluation of coupling terms between excited state-specific multi-configurational 
wave functions\cite{Shea2018,Hardikar2020,Tran2019,Tran2020,Kossoski2022} 
that are prohibitively expensive using the generalized Slater--Condon rules.
Ultimately, the generality of this framework, and the availability of the open-source \LibGNME{} 
implementation,\cite{LibGNME} will cataylse a new phase of development
in nonorthogonal electronic structure theory.

\section*{Acknowledgements}
H.G.A.B.\ was supported by New College, Oxford through the Astor Junior Research Fellowship.
The author thanks Nicholas Lee and Alex Thom for helpful feedback during the preparation of this manuscript,
and Rebecca Lloyd for careful proof-reading.

\section*{Data availability}
The data that supports the findings of this study are available within the article.

\section*{References}
\bibliography{gnme}

\end{document}